\documentclass{aa}
\usepackage{epsfig}

\newcommand{\vek}[1]{\mbox{\boldmath$#1$}}

\begin{document}

\title{Electron beam -- plasma system with the return current\\ and directivity of its X-ray emission}

\titlerunning{Beam-plasma system and its X-ray directivity}

\author{M. Karlick\'y \and J. Ka\v{s}parov\'a}

\offprints{Karlick\'y}

\institute{Astronomical Institute of the Academy of Sciences of the Czech
Republic, 25165 Ond\v{r}ejov, Czech Republic, e-mail: karlicky@asu.cas.cz,
kasparov@asu.cas.cz}


\date{Received  / Accepted }

  \abstract
   {}
   {An evolution of the electron distribution function in the
   beam-plasma system with the return current
   is computed numerically for different parameters. The
   X-ray bremsstrahlung corresponding to
    such an electron distribution is calculated and
   the directivity of the X-ray emission is studied.}
   {For computations of the electron distribution functions we used
   a 3-D particle-in-cell electromagnetic code. The directivity of the X-ray emission
   was calculated using the angle-dependent electron-ion bremsstrahlung cross-section.}
{It was found that the resulting electron distribution function depends on the
magnetic field assumed along the electron beam propagation direction. For small
magnetic fields the electron distribution function becomes broad in the
direction perpendicular to the beam propagation due to the Weibel instability
and the return current is formed by the electrons in a broad and shifted bulk
of the distribution. On the other hand, for stronger magnetic fields the
distribution is more extended in the beam-propagation direction and the return
current is formed by the electrons in the extended distribution tail. In all
cases, the anisotropy of the electron distribution decreases rapidly due to
fast collisionless processes. However, the magnetic field reduces this
anisotropy decrease. The X-ray directivity shows the same trend and it is
always closer to the isotropic case than that in a simple beaming model.}
 {}


\maketitle

\section{Introduction}

It is commonly believed that the hard X-ray emission in solar flares is
produced by the bremsstrahlung process of energetic electrons in dense layers
of the solar atmosphere (Brown 1971; Tandberg-Hanssen \& Emslie 1988).

It is also known that up to now this scenario has several unresolved drawbacks
as summarized in the paper by Brown et al. (1990). For example, the
bremsstrahlung mechanism generating the hard X-ray bursts is of a very low
efficiency and therefore huge electron beam fluxes $E_\mathrm{F}$ = 10$^{9}$ -
10$^{12}$ ergs s$^{-1}$ cm$^{-2}$ are required for an explanation of the
observed X-ray fluxes (Hoyng et al. 1978). It means that at the acceleration
site in the low corona with a relatively low density ($n_\mathrm{e} \sim$
10$^9$ cm$^{-3}$), a substantial part of all plasma electrons needs to be
accelerated. Furthermore, these electron beams represent huge electric currents
that have to be neutralized by the return currents. The return current is a
natural part of any beam-plasma system (van den Oord 1990).

The beam-plasma interaction has been studied for a long time, starting with the
paper by Bohm \& Gross (1949). While the first 1-D models considered the
electrostatic aspects of this interaction (two-stream instability, generation
of Langmuir waves, and quasi-linear relaxation of the beam, see e.g. Melrose
1980, Birdsall \& Langdon 1985, Benz 1993, Karlick\'y 1997 and the references
therein), new 3-D studies include the return current and electromagnetic
effects which lead to many further instabilities (Weibel, filamentation,
oblique, Bell, Buneman, and so on, see Karlick\'y 2009, Bret 2009). (Remark:
The Weibel instability in the sense used here and in the paper by Nishikawa et
al. (2008) is also known as the filamentation instability (Bret 2009).) To
cover all these processes, especially inductive processes neutralizing the
total electric current, in the present study we use a general and fully
self-consistent (basic plasma physics) approach -- a 3-D electromagnetic
particle-in-cell (PIC) modelling.

All the abovementioned processes necessarily modify the electron distribution
function in the flare X-ray source. Moreover, contrary to simple models, which
generally predict high anisotropy of electrons and X-rays, it was found that
the observed hard X-ray directivities are low (e.g. Kane 1983). Furthermore,
Kontar \& Brown (2006) found a low anisotropy of the electron distribution
function in the X-ray source by separating the reflected X-ray emission from
the direct one. They concluded that the conventional solar flare models with
downward beaming are excluded.

In the present paper we want to demonstrate the importance of the
abovementioned processes on the evolution of the beam-plasma system with the
return current. Our aim is to show their effects on the anisotropy of the
electron distribution function in this system and thus on the directivity of
the corresponding X-ray emission. Using the 3-D electromagnetic PIC model, for
the first time in the study of X-ray directivity, we compute the evolution of
the beam-plasma system with the return current depending on the magnetic field
in the beam propagation direction. Then, assuming that the resulting electron
distribution functions generate X-ray bremsstrahlung, we calculate the
directivity of the associated X-ray emission. (For a detailed analysis of the
instabilities and waves produced in the studied beam-plasma system, see
Karlick\'y et al. 2008, Karlick\'y 2009, Karlick\'y and B\'arta 2009.)

The layout of the paper is as follows: In Section 2 we outline our model. The
results of computations of the electron distribution functions with the return
current are shown in Section 3. In Section 4 we present the corresponding X-ray
directivities. Finally, in Section 5 the results are discussed and conclusions
given.

\section{Model}

\begin{table}[t]
\begin{minipage}[t]{\columnwidth}
\caption{Model parameters.}
\label{tab4}
\centering
\renewcommand{\footnoterule}{}
\begin{tabular}{ccccc}
\hline
Model & $m_\mathrm{i}$/$m_\mathrm{e}$  & $n_\mathrm{b}$/$n_\mathrm{e}$  & $v_\mathrm{b}/c$ & $\omega_\mathrm{ce}/\omega_\mathrm{pe}$ \\
\hline
A & 16 & 1/8 & 0.666 & 0.0  \\
B & 16 & 1/8 & 0.666 & 0.1 \\
C & 16 & 1/8 & 0.666 & 0.5 \\
D & 16 & 1/8 & 0.666 & 0.7 \\
E & 16 & 1/8 & 0.666 & 1.0 \\
F & 16 & 1/8 & 0.666 & 1.3 \\
G &  1 & 1/8 & 0.666 & 0.0 \\
H &  1 & 1/8 & 0.666 & 1.3 \\
I & 100 & 1/8 & 0.666 & 0.0 \\
J & 100 & 1/8 & 0.666 & 1.3 \\
K & 16 & 1/8 & 0.333 & 0.0 \\
L & 16 & 1/8 & 0.333 & 1.3 \\
M & 16 & 1/40 & 0.666 & 0.0 \\
N & 16 & 1/8 & 0.234\footnote{mean velocity of the power-law
beam distribution} & 0.0 \\
O & 16 & 1/8 & 0.234\footnote{mean velocity of the power-law
beam distribution} & 1.3 \\
\hline
\end{tabular}
\end{minipage}
\end{table}

For our study we used a 3-D (3 spatial and 3 velocity components) relativistic
electromagnetic PIC code (Buneman 1993). The system sizes are $L_x$ =
45$\Delta$, $L_y$ = 45$\Delta$, and $L_z$ = 600$\Delta$ (where $\Delta$ is the
grid size).

For a basic set of models we initiated a spatially homogeneous electron-proton
plasma with the proton-electron mass ratio $m_\mathrm{p}/m_\mathrm{e}$=16
(Models A-F, and K-O in Table 1). This is unrealistic and it was chosen to
shorten the proton skin depth and computations. Nevertheless, the ratio is
still sufficient to well separate the dynamics of electrons and protons. For
comparison we added models with the mass ratio $m_\mathrm{p}/m_\mathrm{e}$=1
and 100 (Models G-J in Table 1). The electron thermal velocity is
$v_{T\mathrm{e}}$ = 0.06 $c$ (the corresponding temperature is $T_\mathrm{e}$ =
21.4 MK), where $c$ is the speed of light. In all models, 160 electrons and 160
protons per cube grid were used. The plasma frequency is $\omega_\mathrm{pe}$ =
0.05 and the electron Debye length is $\lambda_\mathrm{D}$ = 0.6 $\Delta$. In
the models with the proton-electron mass ratio $m_\mathrm{p}/m_\mathrm{e}$=16,
the electron and proton skin depths are $\lambda_\mathrm{ce}$ = 10 $\Delta$ and
$\lambda_\mathrm{ci}$ = 40 $\Delta$, respectively.

\begin{table}
\begin{minipage}[t]{\columnwidth}
\caption{The real spatial and time scales
as a function of the chosen plasma density $n_\mathrm{e}$.}
\label{catalog}
\centering
\renewcommand{\footnoterule}{}  
\begin{tabular}{ccccc}
\hline
$n_\mathrm{e}$ & $\omega_\mathrm{pe}$ & t = 200/$\omega_\mathrm{pe}$  & $\lambda_\mathrm{D}$ & 1/$\nu_0$ \\
(cm$^{-3}$) & (s$^{-1}$) & (s) & (cm) & (s) \\
\hline
10$^8$ & 5.64 $\times$ 10$^8$ & 3.55 $\times$ 10$^{-7}$ & 3.19 & 3.61 \\
10$^9$ & 1.78 $\times$ 10$^{9}$ & 1.12 $\times$ 10$^{-7}$ & 1.01 & 0.36 \\
10$^{10}$ & 5.64 $\times$ 10$^{9}$ & 3.55 $\times$ 10$^{-8}$ & 0.32 & 0.03 \\
10$^{11}$ & 1.78 $\times$ 10$^{10}$ & 1.12 $\times$ 10$^{-8}$ & 0.10 & 0.003 \\
\hline
\end{tabular}
\end{minipage}
\end{table}

Then, we included one monoenergetic beam homogeneous throughout the numerical
box (see Models A-M). Note that due to the physical and numerical simplicity
and the propagation effect in which faster electrons escape from the slower
ones, in most cases we consider monoenergetic electron beams, although in the
interpretation of solar flare hard X-rays, the power-law distributions are
used. The power-law distributions are derived as mean distributions over the
whole X-ray source for much longer timescales than those considered in the
present study. In much smaller flare volumes and on much shorter timescales,
the monoenergetic beam is a reasonable choice. Nevertheless, in Models N and O
we added computations with the beam having a power-law distribution function.
To show effects of instabilities distinctly we chose its power-law index (in
the velocity space) as 1.5, and the low-velocity cutoff of 0.09 $c$.

\begin{figure}[b]
  \begin{center}
    \epsfig{file=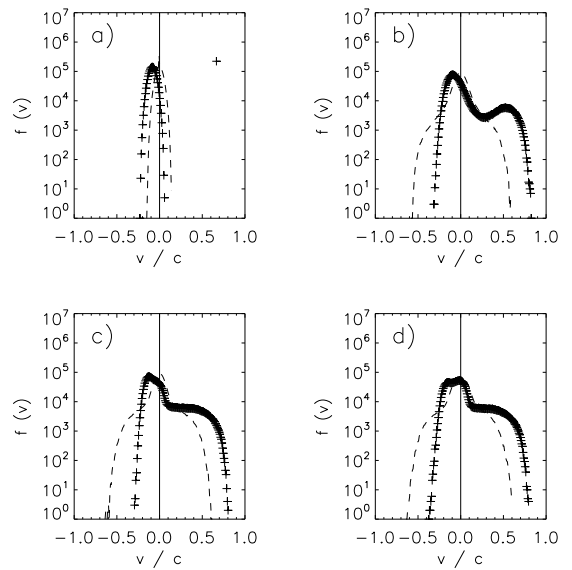, width=8 cm}
    \end{center}
    \caption{The electron distribution functions in Model B
    at four different times:
    at the initial state
 (a), at $\omega_\mathrm{pe} t$ = 40 (b), at  $\omega_\mathrm{pe} t$ = 100
 (c), and $\omega_\mathrm{pe} t$ = 200 (d).
 Crosses correspond to $f(v_{z})$, dotted and dashed lines display $f(v_x)$ and
$f(v_y)$, respectively.
  Note that $f(v_x)$ and $f(v_y)$ overlap.
  The single cross in the part a) at $v/c$ = 0.666
  denotes the monoenergetic electron beam.}
  \label{fig1}
\end{figure}

To keep the total current zero in these models in the initial states, we
shifted the background plasma electrons in the velocity space (i.e. we
initiated the return current) according to the relation $v_\mathrm{d} = -
v_\mathrm{b} n_\mathrm{b}/n_\mathrm{e}$, where $v_\mathrm{b}$ is the velocity
of the electron beam, $n_\mathrm{b}$ and $n_\mathrm{e}$ are the beam and
background plasma densities (for this type of initiation see Niemiec et al.
2008). The beam velocity was chosen to be $v_\mathrm{b}/c$ =  0.666 or 0.333
(in the $z$ direction), see Table 1. The ratio of the beam and plasma densities
was taken as $n_\mathrm{b}/n_\mathrm{e}$ = 1/8 (Models A-L and N-O), and
$n_\mathrm{b}/n_\mathrm{e}$ = 1/40 (Model M).

\begin{figure*}[t]
  \begin{center}
        \epsfig{file=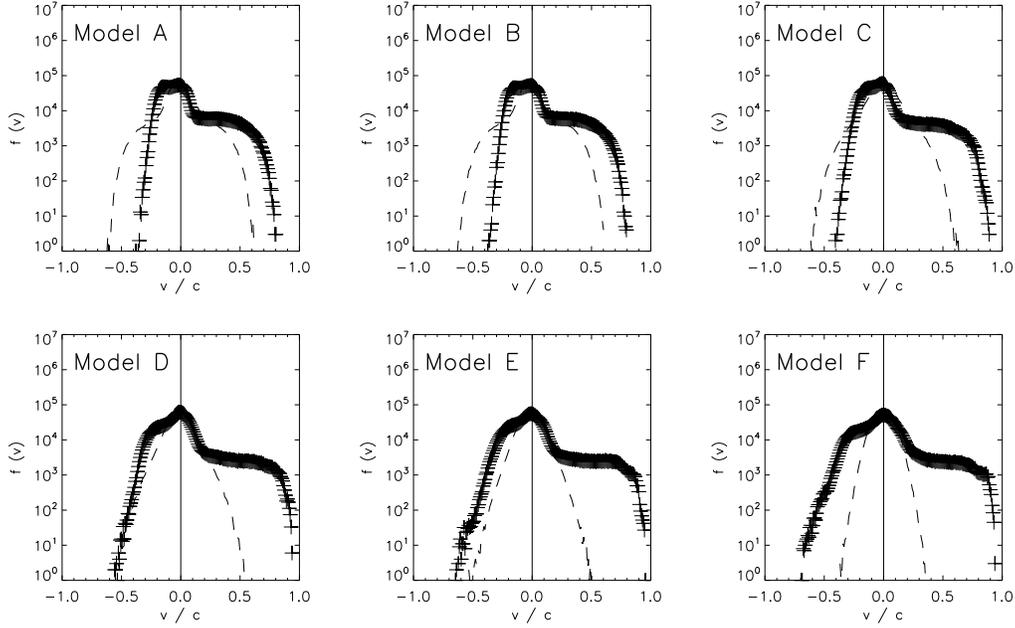, width=14 cm}
     \end{center}
    \caption{The electron distribution
    functions
    at $\omega_\mathrm{pe} t$ = 200
    as a function of the magnetic field
    in Models A-F with $\omega_\mathrm{ce}/\omega_\mathrm{pe}$ =
    0.0, 0.1, 0.5, 0.7, 1.0, and 1.3, respectively.
    Notation is the same as in Fig.~\ref{fig1}.}
  \label{fig2}
\end{figure*}

Because computations in the PIC models are dimensionless, the results are valid
for a broad range of plasma densities.  The real time and spatial scales are
given by specifying the plasma density. Table 2 summarizes temporal and spatial
scales (the interval of computations t = 200/$\omega_\mathrm{pe}$ and the Debye
length) for the plasma densities in the 10$^8$-10$^{11}$ cm$^{-3}$ range. The
processes under study are very fast. The collisional processes are much longer,
see the collisional free time (1/$\nu_0$) in Table 2. The numerical system size
is small (45$\Delta$ x 45 $\Delta$ x 600 $\Delta$ = 75 $\lambda_\mathrm{D}$ x
75 $\lambda_\mathrm{D}$ x 1000 $\lambda_\mathrm{D}$, i.e. for the plasma
density e.g. $n_\mathrm{e}$ = 10$^{9}$ cm$^{-3}$ it gives 76 cm x 76 cm x 1010
cm). Since the periodic boundary conditions are used, in reality the studied
problem is infinite in space.

The beam density and the corresponding beam energy flux is given by the chosen
plasma density $n_\mathrm{e}$, $n_\mathrm{b}/n_\mathrm{e}$ = (1/8 and 1/40),
and the beam velocities (see Table 1). For example, for $n_\mathrm{e}$ = 10$^9$
cm$^{-3}$, n$_b$/n$_e$ = 1/8, and v$_b$ = 0.666 c, the beam density
$n_\mathrm{b}$ = 1.25 $\times$ 10$^8$ cm$^{-3}$ and the beam energy flux
$E_\mathrm{flux}$ = 4.55 $\times$ 10$^{11}$ ergs s$^{-1}$ cm$^{-3}$.

Because we want to study the influence of the magnetic field, in the models we
consider several values of the ratio of the electron-cyclotron and
electron-plasma frequencies ($\omega_\mathrm{ce}/\omega_\mathrm{pe}$ = 0.0,
0.1, 0.5, 0.7, 1.0, and 1.3 ~-- see Table 1). Note that in the space close to
the flare acceleration site in the low corona there is plasma of relatively low
density. Thus, for the huge electron beam fluxes required for an explanation of
the observed X-ray bursts, such high ratios of $n_\mathrm{b}/n_\mathrm{e}$ are
needed. In all models, the periodic boundary conditions were used.

\section{Results of 3-D PIC simulations}

As an illustration of the time evolution of the electron distribution function
in the beam-plasma system with the return current, Fig.~1 shows this evolution
for Model B.  As can be seen, due to the two-stream instability (Michailovskij
1975), a plateau of the distribution function $f(v_z)$ (in the beam propagation
direction) on the beam side is formed. Moreover, some small part of the
electrons even increased their energy due to their interaction with generated
Langmuir waves. Simultaneously, the distribution functions $f(v_x)$ and
$f(v_y)$, i.e. the distribution functions in the directions perpendicular  to
that of the beam propagation, are strongly heated. This is due to the Weibel
instability (1959) (see also Nishikawa at al. 2006).

To demonstrate how the magnetic field influences the resulting electron
distribution function, Fig.~2 presents the distribution functions for six
values of the ratio of the electron-cyclotron and electron-plasma frequencies
($\omega_\mathrm{ce}/\omega_\mathrm{pe}$ = 0.0, 0.1,0.5, 0.7, 1.0, and 1.3 -
Models A-F, Table 1). It is evident that with the increase of the ratio
$\omega_\mathrm{ce}/\omega_\mathrm{pe}$, the role of the Weibel instability is
more and more reduced, the distribution functions in the direction
perpendicular to the beam propagation $f(v_x)$ and $f(v_y)$ are less heated. On
the other hand, the problem of the return current formation becomes more and
more one-dimensional and a more extended tail on the return current side is
formed (compare Model A and F in Fig.~2, see also Karlick\'y et al. 2008;
Karlick\'y 2009). In Fig.~3 the same results are expressed in terms of the
electron distribution functions depending on the electron energies. Although
this type of description is more common in flare research, the distribution
functions in
\begin{figure*}[t]
  \begin{center}
    \epsfig{file=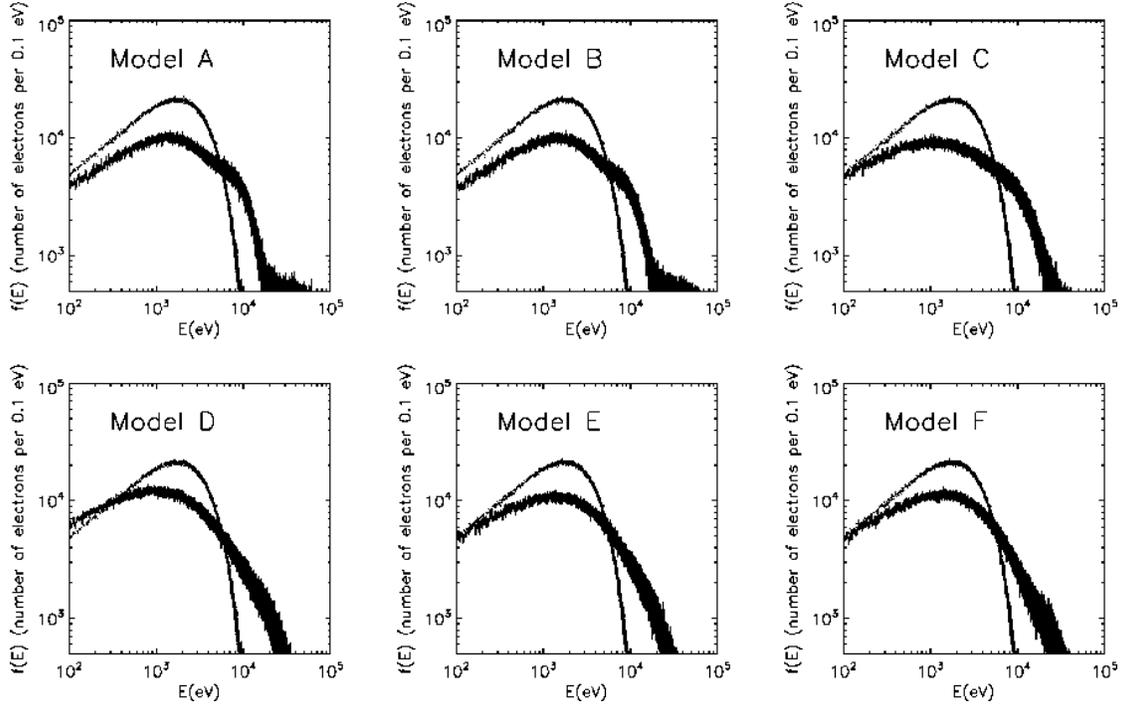, width=0.8\textwidth}
        \end{center}
    \caption{The electron distribution
    functions in electron energies (thick lines)
    at $\omega_\mathrm{pe} t$ = 200
    as a function of the magnetic field
    in Models A-F with $\omega_\mathrm{ce}/\omega_\mathrm{pe}$ =
    0.0, 0.1, 0.5, 0.7, 1.0, and 1.3, respectively. For comparison
    in each panel the initial electron plasma distribution is added (thinner lines).}
  \label{fig3}
\end{figure*}
\begin{figure*}[!t]
  \begin{center}
    \epsfig{file=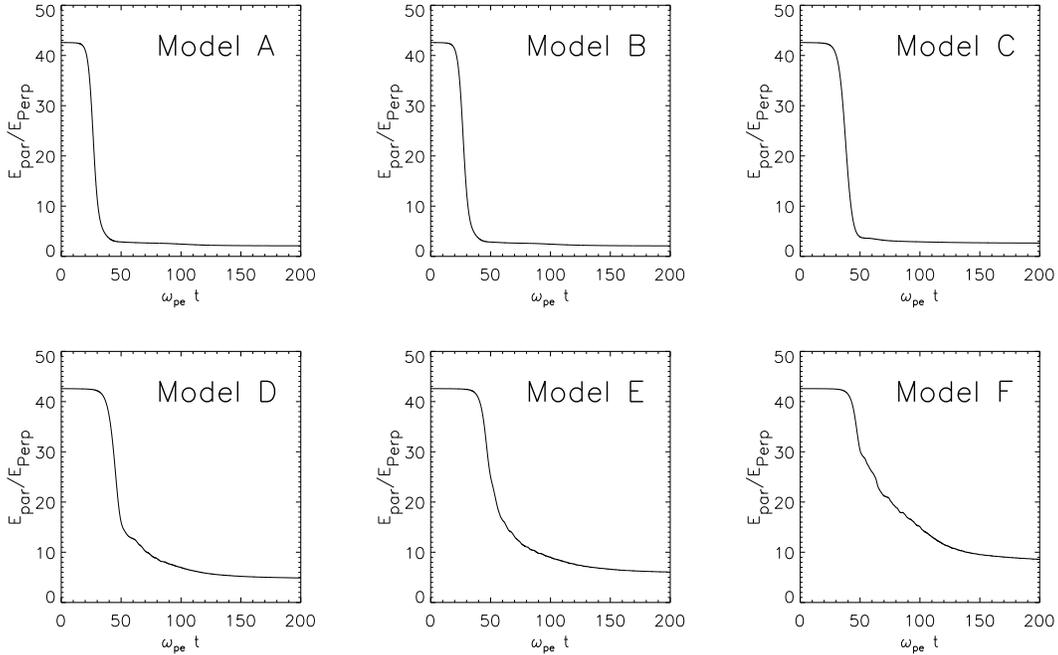, width=0.8\textwidth}
        \end{center}
    \caption{Time evolution of the ratio of the electron kinetic parallel and
    perpendicular energies $E_\mathrm{par}/E_\mathrm{per}$
    as a function of the magnetic field
    in Models A-F with $\omega_\mathrm{ce}/\omega_\mathrm{pe}$ =
    0.0, 0.1, 0.5, 0.7, 1.0, and 1.3, respectively.}
  \label{fig4}
\end{figure*}
velocity space presented in Fig.~2 carry more information than those in Fig.~3
and thus they are more physically relevant in describing the studied processes.
The ratio of the electron kinetic energies in the direction parallel and
perpendicular to that of beam propagation, which expresses the "anisotropy" of
the system, is shown in Fig.~4. The ratio of energies is defined as:
\begin{eqnarray}
\frac{E_\mathrm{par}}{E_\mathrm{perp}} = \frac{\sum_{i=1}^n \frac{1}{2} m_\mathrm{e}
v_{iz}^2}{\sum_{i=1}^n \frac{1}{4} m_\mathrm{e} (v_{ix}^2 + v_{iy}^2)},
\end{eqnarray}
where $n$ is the number of electrons in the whole numerical box.  As can be
seen in Fig. 4, the collisionless (wave-particle) processes very rapidly
decrease the "anisotropy" on time scales shorter than $\omega_\mathrm{pe}
t\approx 50$. This process is faster and more efficient for lower magnetic
fields. While the ending ratio is $E_\mathrm{par}/E_\mathrm{perp} \approx$ 9
for Model F ($\omega_\mathrm{ce}/\omega_\mathrm{pe}$ = 1.3), in Model A
($\omega_\mathrm{ce}/\omega_\mathrm{pe}$ = 0.0) this ratio is only
$E_\mathrm{par}/E_\mathrm{perp} \approx$ 2.

In Fig.~5 a comparison of models with three different mass ratios
($m_\mathrm{p}/m_\mathrm{e}$ = 1, 16, 100) and two values of the ratio
$\omega_\mathrm{ce}/\omega_\mathrm{pe}$ (0.0 and 1.3) is made. While in the
cases with $m_\mathrm{p}/m_\mathrm{e}$ = 1 (the electron-positron plasma) the
strong heating of the distribution functions $f(v_x)$ and $f(v_y)$ can be seen
even for the strong magnetic field ($\omega_\mathrm{ce}/\omega_\mathrm{pe}$ =
1.3), for the proton-electron plasma the resulting $f(v_x)$ and $f(v_y)$ for
$m_\mathrm{p}/m_\mathrm{e}$ = 16 and 100 do not differ significantly. Note that
in the model with $m_\mathrm{p}/m_\mathrm{e}$ = 100 the proton skin depth is
greater than the system sizes $L_x$ and $L_y$.

\begin{figure}[t]
  \begin{center}
        \epsfig{file=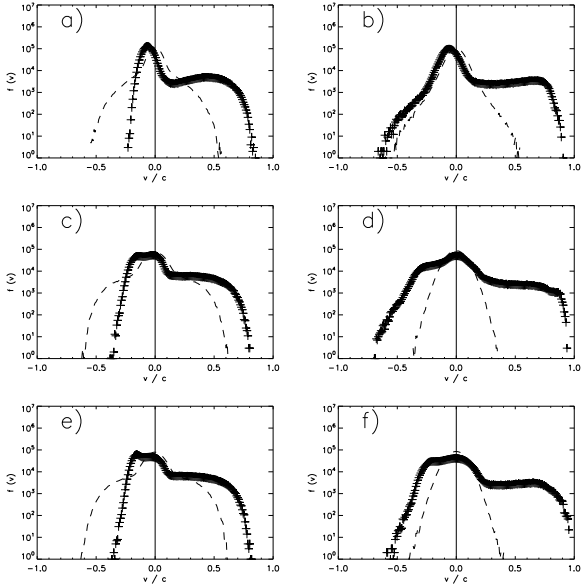, width=8 cm}
            \end{center}
    \caption{The electron distribution
    functions
    at $\omega_\mathrm{pe} t$ = 200
    as a function of the mass ratio:
    $m_\mathrm{i}/m_\mathrm{e}$ = 1 -- two upper plots,
    $m_\mathrm{i}/m_\mathrm{e}$ = 16 -- two middle plots, and
    $m_\mathrm{i}/m_\mathrm{e}$ = 100 -- two bottom plots 
    for two values of $\omega_\mathrm{ce}/\omega_\mathrm{pe}$ = 0.0
    (left column) and 1.3 (right column).
    Notation is the same as in Fig.~\ref{fig1}.}
  \label{fig5}
\end{figure}

We also compared the evolution of the electron distribution functions in Models
A and F with Models K and L, i.e. the models with a lower initial beam velocity
($v_\mathrm{b}/c$ = 0.333). We found that only the extent of the return-current
tail in Model L is shorter than that in Model F. It is a natural consequence of
the greater beam velocity in Model F than in Model L. Furthermore, it was found
that Model M gave qualitatively the same results as Model A.

In Figs. 6 and 7 the electron distribution functions in Models N and O, i.e. in
the models with the power-law beam and with two different ratio of
electron-cyclotron and electron-plasma frequencies
($\omega_\mathrm{ce}/\omega_\mathrm{pe}$ = 0.0 and 1.3) are shown. Because
these models are not subject  to the bump-on-tail instability there are no
significant changes in the distribution $f(v_z)$ on the beam distribution side.
On the other hand, the Weibel instability plays its role, especially in the
case without the magnetic field (Model N). Once again, in Model N the plasma is
heated in the direction perpendicular to that of beam propagation, whereas in
Model O, the return current is formed by the extended distribution tail.

\begin{figure}[t]
  \begin{center}
    \epsfig{file=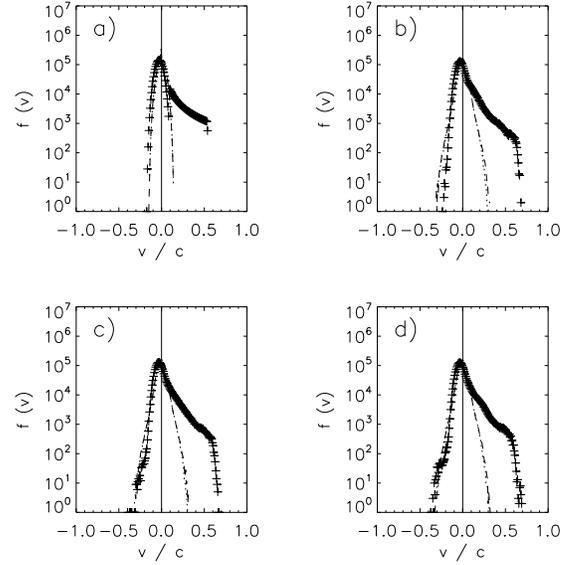, width=8 cm}
        \end{center}
    \caption{The electron distribution functions in Model N with the power-law beam
    at four different times: at the initial state (a),
    at $\omega_\mathrm{pe} t$ = 40 (b), at  $\omega_\mathrm{pe} t$ = 100 (c),
    and $\omega_\mathrm{pe} t$ = 200 (d).
    Notation is the same as in Fig.~\ref{fig1}.}
  \label{fig6}
\end{figure}

\begin{figure}[t]
  \begin{center}
    \epsfig{file=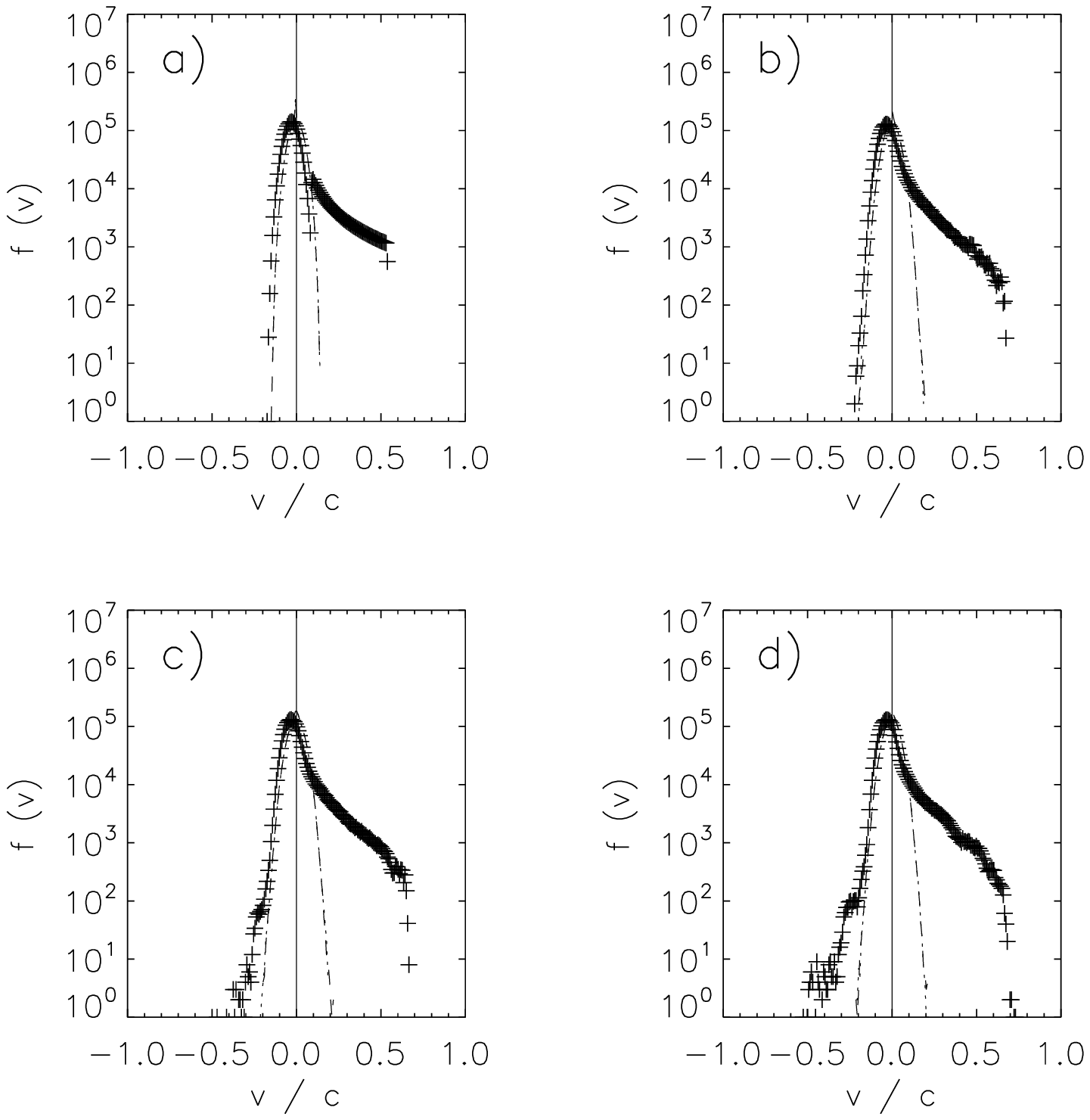, width=8 cm}
        \end{center}
    \caption{The electron distribution functions in Model O
      with the power-law beam
      at four different times: at the initial state (a),
      at $\omega_\mathrm{pe} t$ = 40 (b), at  $\omega_\mathrm{pe} t$ = 100 (c),
      and $\omega_\mathrm{pe} t$ = 200 (d).
      Notation is the same as in Fig.~\ref{fig1}.}
  \label{fig7}
\end{figure}

\section{Directivity of X-ray emission}
\begin{figure*}[t]
  \begin{center}
    \epsfig{file=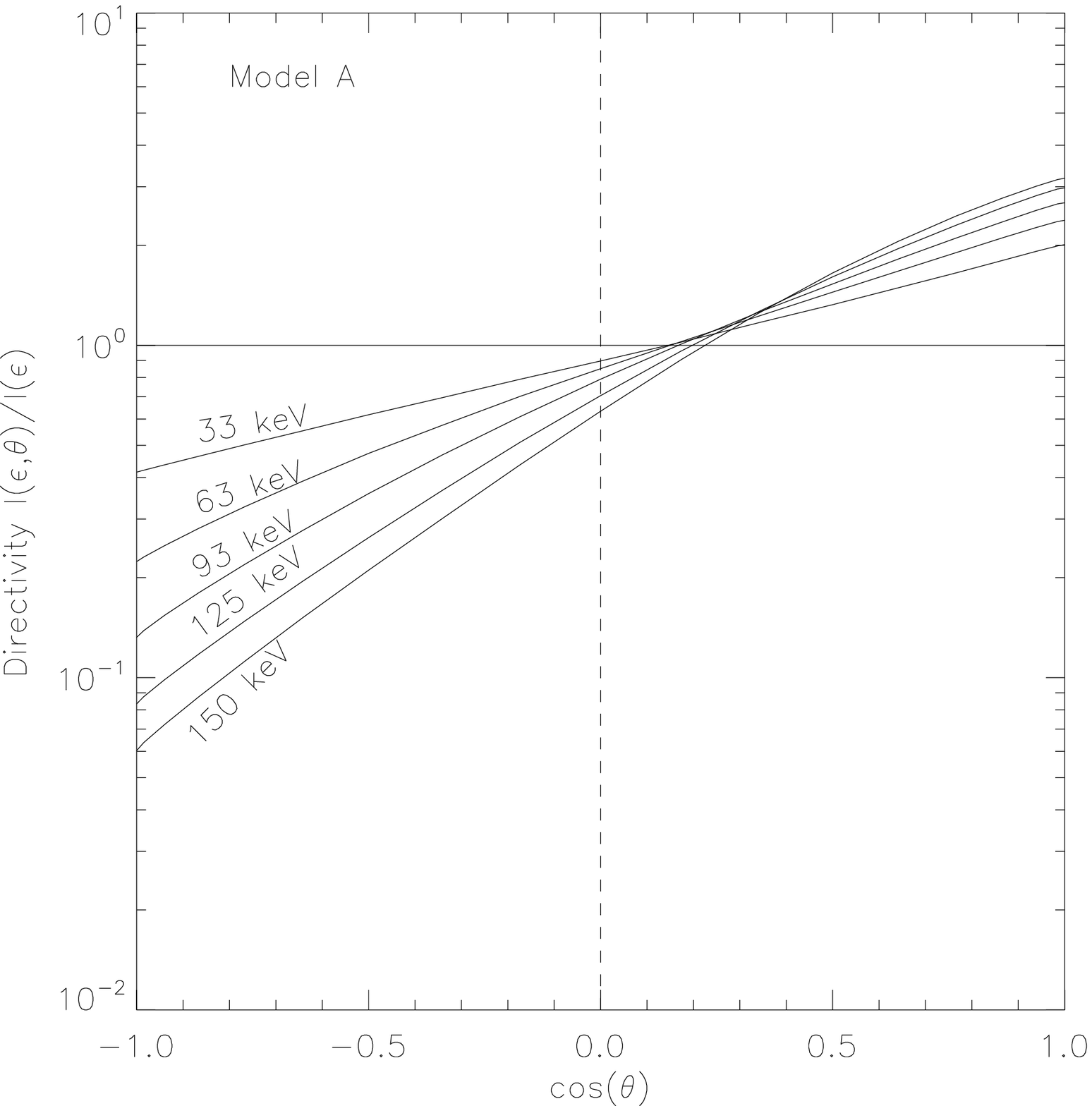, width=6 cm}
    \epsfig{file=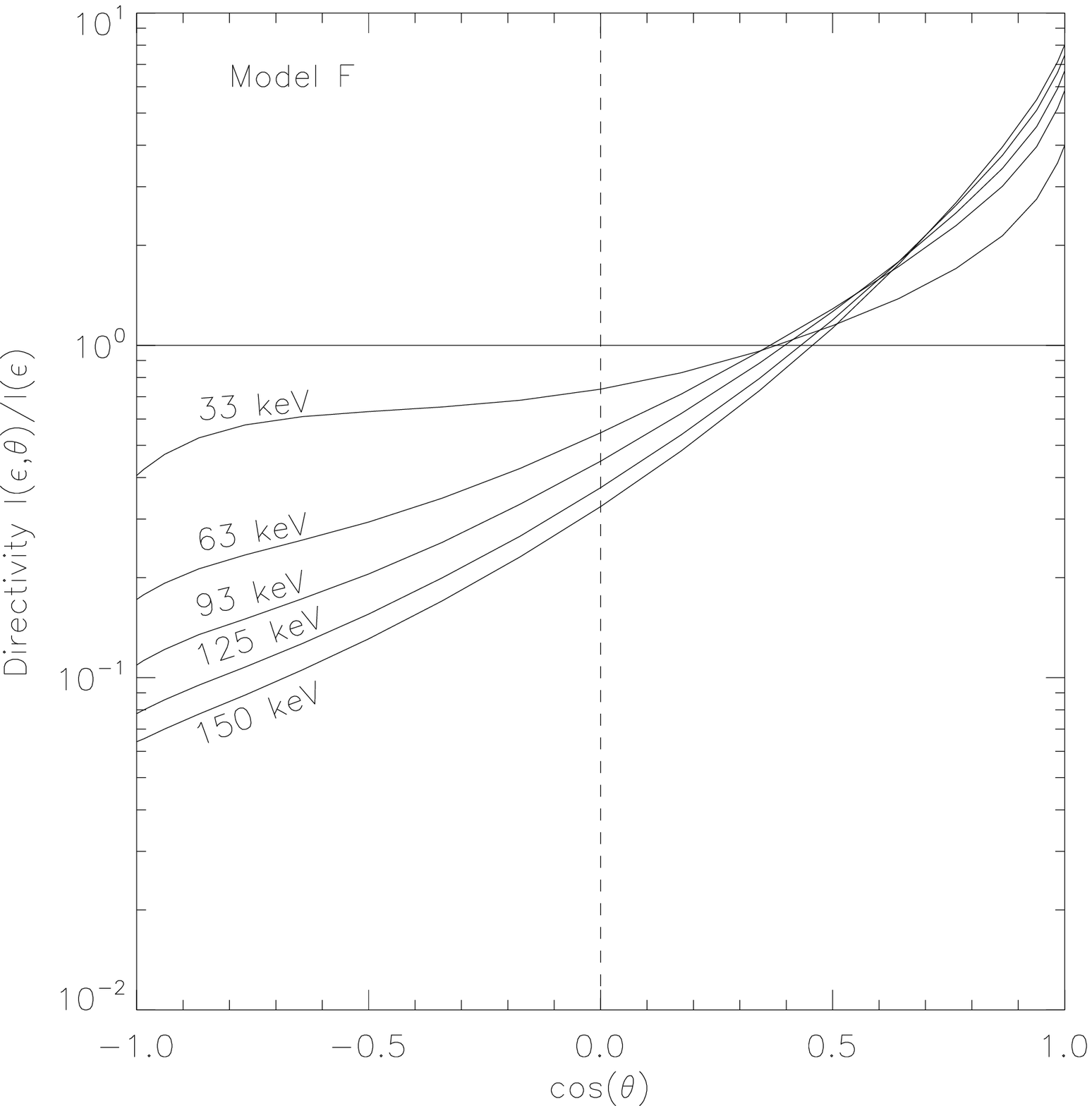, width=6 cm}
    \epsfig{file=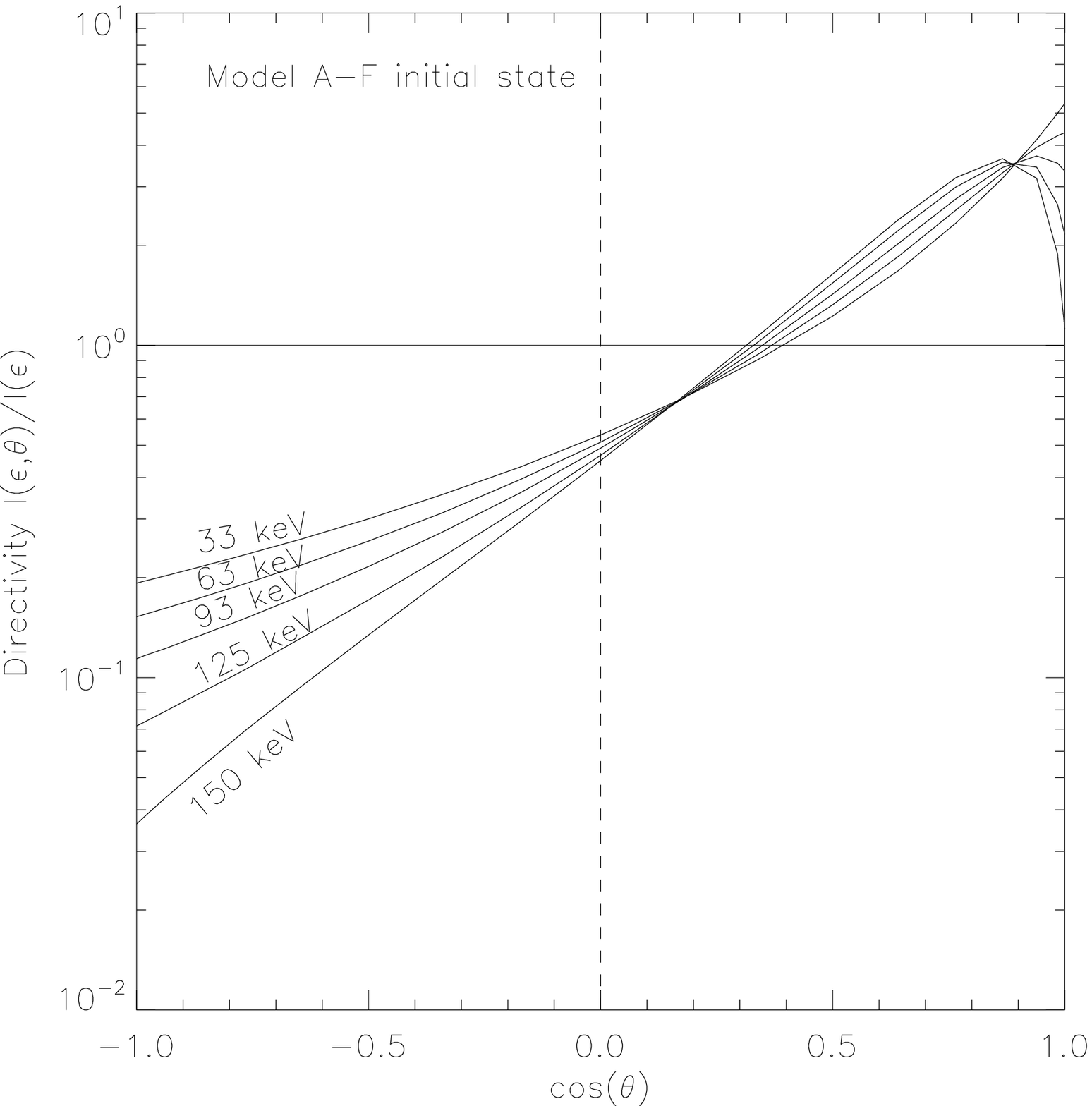, width=6 cm}
    \end{center}
  \caption{The X-ray directivity in several energies for $f(\vek{v})$
  corresponding to Models A and F
  at $\omega_\mathrm{pe}t=200$ and the X-ray directivity
in the initial state for Models A-F (the case of simple beaming). The
horizontal solid line represents the isotropic case,
 the dashed vertical line denotes the viewing angle for a limb source.}
  \label{fig8}
\end{figure*}
\begin{figure*}[t]
  \begin{center}
    \epsfig{file=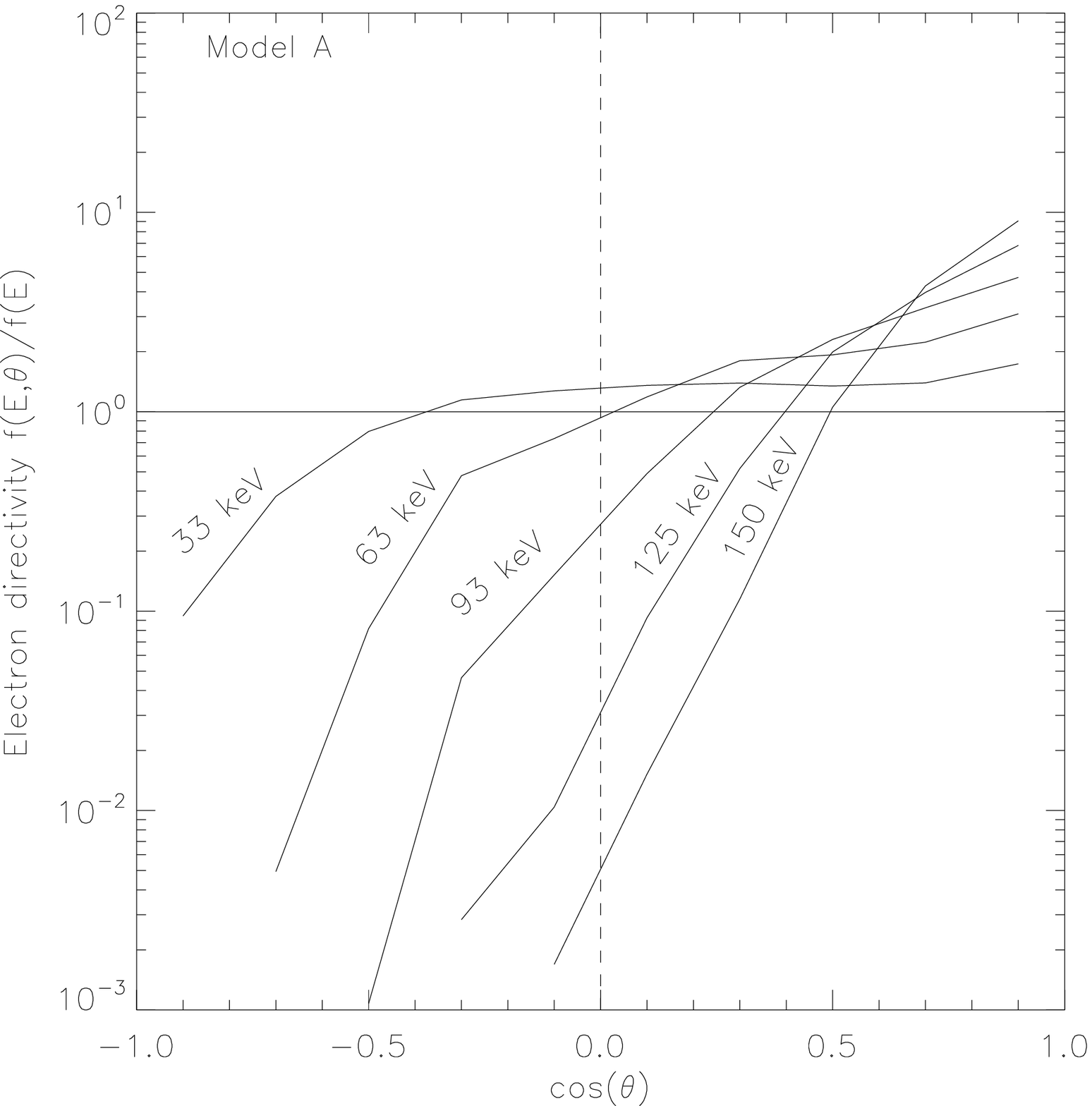, width=6 cm}
    \epsfig{file=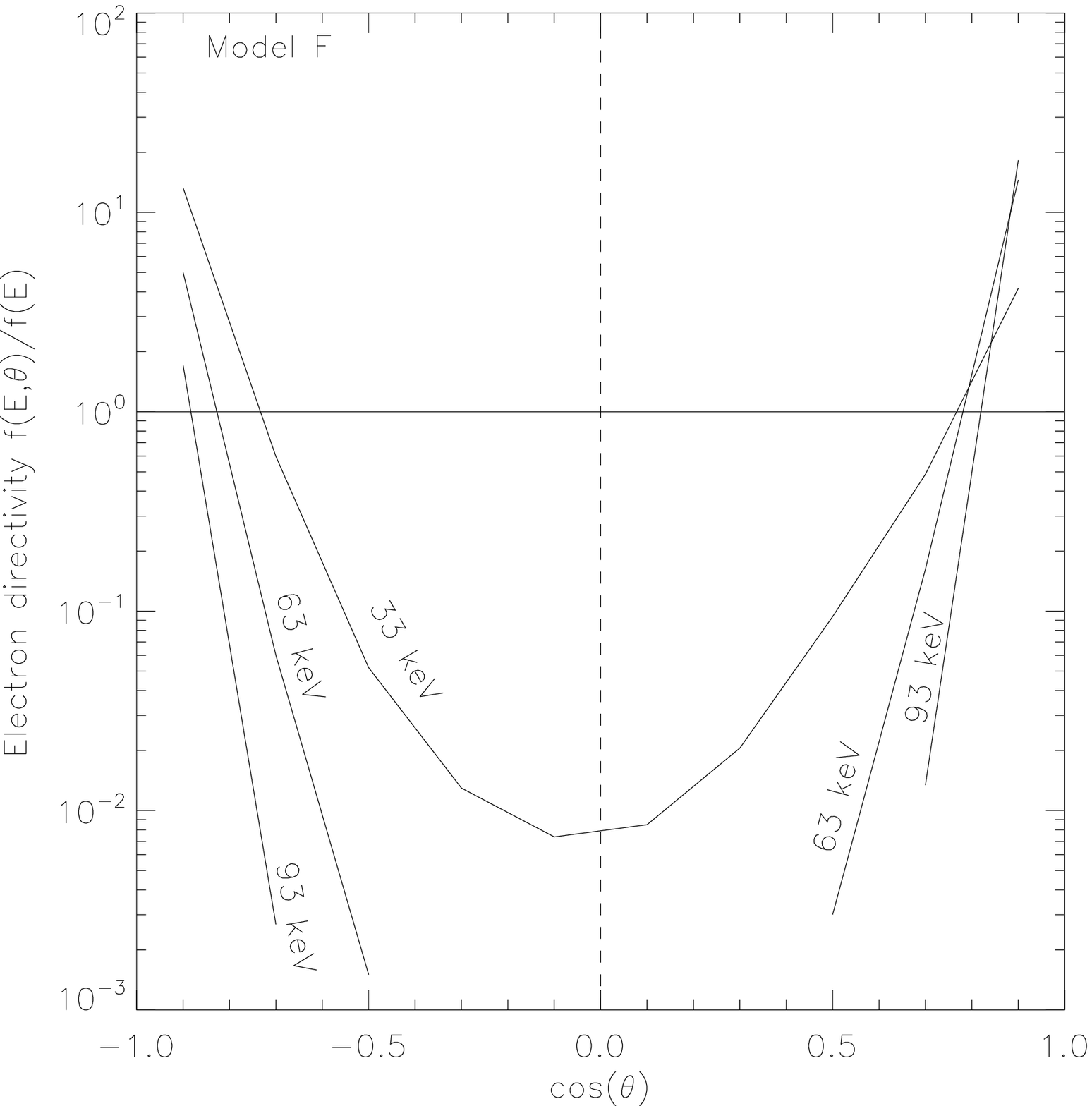, width=6 cm}
    \end{center}
  \caption{The electron directivity in several energies
  for Models A and F at $\omega_\mathrm{pe}t=200$.
The horizontal solid line represents the isotropic case,
 the dashed vertical line denotes the viewing angle for a limb source.
 The corresponding X-ray directivities are shown in Fig.~\ref{fig8}.}
  \label{fig9}
\end{figure*}
Knowing the electron distribution function $f(\vek{v})$, an instantaneous X-ray
bremsstrahlung, i.e. the so-called thin-target emission (e.g. Brown et al.,
2003) can be calculated. To account for  the anisotropy of $f(\vek{v})$, we
considered the angle-dependent electron-ion bremsstrahlung cross-section
$Q(\epsilon,E,\Theta)$ differential in the electron energy $E$ and the solid
angle of the incoming electron, where $\epsilon$ is the photon energy and
$\Theta$ is the angle between the electron pre-collision velocity and direction
of the photon emission (Gluckstern \& Hull, 1953). We used the expression for
$Q(\epsilon, E,\theta)$ given in Appendix of Massone et al. (2004), which
includes the Elwert (1939) Coulomb correction.  The cross-section was evaluated
using \verb(hsi_reg_ge_angle_cross.pro( available in the Solar Software.

Figure~\ref{fig8} shows the X-ray directivity, i.e. the ratio of the
angle-dependent $I(\epsilon, \theta)$ to integral photon spectrum
$I(\epsilon)=1/4\pi\int_{\Omega} I(\epsilon,\theta,\phi)\ \mathrm{d}\Omega$,
where $\theta$ and $\phi$ is the polar and azimuthal angle, respectively,
$\Omega$ is the solid angle. The $z$-axis of the coordinate system is chosen to
be along the beam propagation direction. Note that due to axial symmetry of the
problem around the $z$-axis, the photon spectrum $I(\epsilon, \theta,\phi)$ is
also independent of $\phi$, so $I(\epsilon, \theta)=I(\epsilon,\theta,\phi)$.
Assuming that the beam propagates along the local normal line towards the
photosphere, Fig.~\ref{fig8} displays a variation of the X-ray directivity
observed from different viewing angles: the cases with $\cos\theta=1$ and
$\cos\theta= -1$ correspond to the forward (the direction to the photosphere)
and backward (the direction to the Earth's observer when the X-ray source is at
the disc centre) emissions, while the case with $\cos\theta=0$ denotes the
emission in the perpendicular direction (the X-ray source placed on the solar
limb).

The behaviour of the X-ray directivity is closely related to the corresponding
electron distribution. Comparing Model A and F at the time $\omega_\mathrm{pe}
t$ = 200 with Models A-F in the initial state (i.e. the case with a simple
beaming) in Fig.~\ref{fig8}, it can be seen that values of the directivity,
especially in the backward direction, become closer to the value 1 (the
isotropic case). Therefore, the global directivity decreased during the
evolution of the electron distribution. Furthermore, we can see that the
directivity values for $\cos\theta=0$ in Model A  are closer to the isotropic
case than those in Model F. This is due to the strong heating of the plasma in
the direction perpendicular to the beam propagation and it is caused by the
Weibel instability in Model A (the case with zero magnetic field).

We also defined the electron directivity $f(E,\theta)/f(E)$, similarly to the
X-ray one. Models A and F at time $\omega_\mathrm{pe} t$ = 200 are presented in
Fig.~\ref{fig9} and show in another way the electron distribution
characteristics discussed above in Section~3, Fig.~\ref{fig2}. Comparing these
electron directivities, we can see that they differ more distinctly than the
corresponding X-ray directivities (Fig.~\ref{fig8}). Such a difference is
caused by the strong smoothing effect of the bremsstrahlung cross-section.

We also calculated the X-ray directivities for Models K-L and N-O. They show
the same changes as follows from the comparison of plots in Fig. 8, but these
changes are less pronounced due to smaller changes of the $f(\vec v)$
anisotropy in these models.

\section{Discussion and conclusions}

Varying the ratio of electron-cyclotron and electron-plasma frequencies
$\omega_\mathrm{ce}/\omega_\mathrm{pe}$, it was found that the magnetic field
influences the evolution of the electron distribution function in electron beam
-- plasma system with a return current. While for small magnetic fields
($\omega_\mathrm{ce}/\omega_\mathrm{pe}$ $\leq$ 0.1) the electron distribution
function becomes broad in the direction perpendicular to the beam propagation
due to the Weibel instability and the return current is formed by the electrons
in a broad and shifted bulk of the distribution, for stronger magnetic fields
($\omega_\mathrm{ce}/\omega_\mathrm{pe}$ $\geq$ 1) the distribution is more
extended in the beam-propagation direction and the return current is formed by
the electrons in an extended distribution tail. Assuming the magnetic field and
electron density as $B$ = 100 G and $n_\mathrm{e}$ = 10$^{11}$ cm$^{-3}$
relevant to solar flares, the ratio of the electron-cyclotron and
electron-plasma frequencies is $\omega_\mathrm{ce}/\omega_\mathrm{pe}$ = 0.1.
In such conditions the Weibel instability plays a role, but it is reduced for a
higher magnetic field. The evolution is influenced also by the two-stream
instability. Besides the formation of the plateau of the electron distribution
on the electron beam side, the simultaneously generated Langmuir waves even
accelerate a small part of the electrons.

The collisionless processes cause a very fast decrease of the ratio of the
electron kinetic parallel and perpendicular (with respect to the beam
propagation direction) energies and lead to a decrease of the "anisotropy" of
the system. Thus, the distribution function rapidly deviates from that with
simple beaming. This can be also expressed by a decrease of the directivity of
the associated X-ray bremsstrahlung emission. This fact agrees with the
statement of Kontar \& Brown (2006) that conventional solar flare models with a
simple downward beaming should be excluded.

An additional aspect of the present study is that the inclusion and physical
necessity of the return current in the beam -- plasma system resolves the
problem of number of electrons  needed for an acceleration of the dense
electron beam in the corona where the density is relatively low. The return
current simply carries the same amount of electrons as in the electron beam
back to the acceleration site. However, the return current does not have the
same distribution function as the initially injected beam.

Variations of the X-ray directivity obtained in our models are of a level
comparable to those in the electron beam propagation models by Langer \&
Petrosian (1977, Fig.~1) and Leach \& Petrosian (1983, Fig.~4). However, there
is an important difference between our model and the models by
\cite{LangerPetrosian77} and \cite{LeachPetrosian83}. We treat only
collisionless processes which were neglected in the previous studies. Due to
the very short time scales in our computations, no effects of longer beam
propagation or collision scattering are included in the electron beam
evolution.

Therefore, the similar level of X-ray directivies suggests that a comparable
level of isotropisation of the electron distribution function caused by the
collisional processes can be produced by the studied wave-particle processes on
much shorter time scales. Moreover, it means that these fast processes should
not be neglected in X-ray directivity studies.

Our study is not aimed at a direct comparison with observations, mainly due to
the large difference between simulated and observationally available time
scales. Nevertheless, the paper by Kontar \& Brown (2006) allows us to compare
our simulations with their derived ratio of downward-to-upward electron
distributions, $F_\mathrm{d}(E)/F_\mathrm{u}(E)$. The comparison reveals an
agreement between inferred $F_\mathrm{d}(E)/F_\mathrm{u}(E)$ and Model F within
the confidence interval up to $\sim$~50~keV. At higher energies, our models
predict a directivity higher than that obtained from observations.

The results presented here could be appropriate for low-density parts of flare
loops where the collisionless processes are dominant. Furthermore, one may
consider them as input into simulations (on much longer time scales) which
treat a propagation of the beam in the environment where Coulomb collisions
play a significant role, such as the transition region and the chromosphere.
Since all these processes (collisionless on long time scales, collisional and
even ionization processes in the background plasma) lead to further
isotropisation of the particle distribution, we speculate that the resulting
electron distribution and X-ray directivity would be much closer to the
isotropic case, as was recently found from X-ray observations (Kontar \& Brown,
2006).

\begin{acknowledgements}
All computations were performed on the parallel computer OCAS (Ond\v{r}ejov
Cluster for Astrophysical Simulations, see http://wave.asu.cas.cz/ocas). This
research was supported by the grant IAA300030701 (GA \v{C}R) and the research
project AV0Z10030501 (Astronomical Institute). The authors thank the referee
for constructive comments that improved the paper.
\end{acknowledgements}

\end{document}